\documentstyle[12pt,twoside,cmp209,epsfig]{article}
\newcommand\bnabla{\mbox{\boldmath $\nabla$}}
\newcommand{\be}{\begin{equation}}
\newcommand{\ee}{\end{equation}}
\newcommand{\<}{\langle}
\renewcommand{\>}{\rangle}
\renewcommand{\r}{{\bf r}}

\newcommand\undemi{{\textstyle\frac 12}}
\newcommand\D{{\cal D}}
\renewcommand\P{{\cal P}}
\newcommand{\centre}[2]{\multicolumn{#1}{l}{#2}}

\def\br{\noalign{\vskip2pt\hrule height1pt\vskip2pt}}

\title[Random Ising model in three dimensions]
	{Random Ising model in three dimensions: theory, 
	experiment and simulation -- a difficult 
	coexistence}
\author[Berche et al]
	{B. Berche\refaddr{*}, P.E. Berche\refaddr{**}, 
	C. Chatelain\refaddr{*}, W. Janke\refaddr{***}
        }
\addresses{
        \addr{*} {Groupe M, Laboratoire de Physique des Mat\'eriaux, }%\\[-1mm]
        {UMR CNRS 7556,}\\%[-1mm]
        {Universit\'e Henri Poincar\'e,  Nancy 1,}\\%[-1mm]
        {F-54506 Vand\oe uvre les Nancy Cedex, France}\\%[1mm]
        \addr{**} {Groupe de Physique des Mat\'eriaux, }%\\[-1mm]
        {UMR CNRS 6634,}\\%[-1mm]
        {Universit\'e de Rouen,}\\%[-1mm]
        {F-76801 Saint Etienne du Rouvray Cedex, France}\\%[1mm]
        \addr{***} {Institut f\"ur Theoretische Physik,}\\%[-1mm]
        {Universit\"at Leipzig,}\\%[-1mm]
        {D-04109 Leipzig, Germany}\\[2mm]
        {\footnotesize\tt berche@lpm.u-nancy.fr}\\%[-1mm]
        {\footnotesize\tt pierre.berche@univ-rouen.fr} \\%[-1mm]
        {\footnotesize\tt chatelai@lpm.u-nancy.fr}\\%[-1mm]
        {\footnotesize\tt wolfhard.janke@itp.uni-leipzig.de}\\[-3mm]
}
\begin{document}

\maketitle

\begin{abstract}
We discuss different approaches for studying the influence of 
disorder in the three-dimensional
Ising model. From the theoretical point of view, renormalisation group calculations 
provide quite accurate
results. Experiments carried out on crystalline mixtures of compounds 
lead to measurements as accurate as three digits on the values of critical exponents. 
Numerically, extensive
Monte Carlo simulations then pretend to be of comparable accuracy. 
Life becomes complicated
when details are compared between the three approaches.
\ {Fluctuation phenomena, random processes, and
Brownian motion}, 
      {64.60.Fr}\ {Equilibrium properties near critical points,
critical exponents},
      {75.10.Hk}\ {Classical spin models}
\end{abstract}

\section{Introduction\label{Sec:Intro}}
Since many years the random Ising model has served as a paradigmatic 
system in which the  
influence of disorder may be studied through different 
techniques. Here we now mention
three of them. The renormalization group (RG) approach, 
experimental measurement and Monte
Carlo simulation. 

RG calculations were
considered quite early~\cite{Khmelnitskii74,Harris74} 
and since then many groups were 
illuminated by skillful RG calculations. Among them, we 
would like to mention the work of 
Folk, Holovatch and Yavors'kii (for recent reviews,
see Refs.~\cite{Folk1,Pelissetto1}). 
On the experimental side, measurements on crystalline mixtures 
of difluoride of different transition metals, e.g., magnetic FeF$_2$ 
substituted with non-magnetic ZnF$_2$,
were performed in the same period over two decades,
gaining in refinement and accuracy (see early results of 
Birgeneau {\em et al.\/} in
Ref.~ \cite{Birgeneau0}, for a review 
see, e.g., Ref.~\cite{Folk1}). 
For the third aspect of simulations, technical progress made by computer 
manufacturers enabled more and more accurate 
simulations (which started for disordered systems in $3D$, e.g., 
with Landau in Ref.~\cite{Landau80}) 
and the study of disordered magnetic 
systems benefited from the development 
of parallel computing. Monte Carlo simulators thus competed in performance
(for a review, see, e.g., Ref.~\cite{Folk1}).

To introduce the subject, we may imagine a conversation between three people
a few decades ago, when the concept of universality was not as clearly
stated as it is nowadays. Imagine a theoretician as the one
who stands up with enthusiasm for this new concept. Let us call him Salviati.
He has an interesting discussion with a good physicist, an experimentalist,
let us say named Sagredo. It is usually considered that the concept of
``numerical experiments'' originates in the FPU problem, 
a numerical study of the thermalization of a chain of atoms, performed by 
Fermi, Pasta and Ulam at Los Alamos~\cite{FPU}.
A third person is thus participating in our conversation. Having no
more character free, let us call him Simplicio -- the simulator!

{%\small
\em\vskip2mm  \noindent In a pub, beginning of the seventies:
\begin{description}
\item{{\sc Sagr.}\ }Dear friends, I would like to report on recent experiments
	that I am conducting at the lab. I produced many samples of
	difluoride of magnetic transition metals, substituting randomly
	different
	amounts of non-magnetic metal, and found very interesting results.
\item{{\sc Salv.}\ }It would be great to compare your results with recent 
        theoretical predictions. Which quantity do you measure?
\item{{\sc Sagr.}\ }Critical temperature, correlation length, 
	susceptibility,\dots
	Of course, the transition temperature decreases when impurities 
	are added, but what looks interesting is the neighbourhood of the
	transition. The singularities
	of some quantities (susceptibility, correlation length) 
	seem to be independent of the impurity concentration.
\item{{\sc Salv.}\ }This is a wonderful observation. 
	It supports the universality
	assumption. You know, from recent RG theory, one expects that
	the free energy density has a singularity in the vicinity of the 
	transition, and that this singularity is  described by some
	critical exponents which are believed to be independent of the
	details of the system under consideration.
\item{{\sc Sagr.}\ } Do you mean that the presence of impurities is a detail? 
	Experimentally
	it is not. It produces an observable decrease of transition 
	temperature.
\item{{\sc Salv.}\ }You are right, critical temperature is not universal and
	disorder is perhaps not a detail, but the precise amount of disorder
	probably does not matter, at least in some range.
\item{{\sc Sagr.}\ }But the singularities that I measured are different from
	those of a pure sample that I have also produced. A colleague of mine
	made similar experiments on Heisenberg-like samples and he
	did not notice any similar modification of the singularities due
	to the introduction of disorder.
\item{{\sc Salv.}\ }Probably that disorder is relevant in your case and not in 
	his case.
\item{{\sc Simpl.}\ }Maybe we could make Monte Carlo simulations, I have access
	to a computer and I have been told that it is not very difficult
	to produce simulations of a disordered Ising model on a cubic lattice.
	I only have to add quenched vacancies in the system and average
	over the disorder realizations.
\item{{\sc Salv.}\ }We may also write an effective $\varphi^4$-theory 
	for the diluted problem
	with a scalar field Ginzburg-Landau-Wilson Hamiltonian 
	and calculate the critical exponents analytically.
\item{{\sc Sagr.}\ }I will measure many other 
	quantities and see what is universal
	and what is not, and we will compare our results.
\end{description}
And this is where the problems occur. In the comparison \dots\vskip2mm
}

We will give in the rest of the paper a short review 
of some recent progress in
studies of the $3D$ disordered Ising model, emphasising the role of
universality and its difficult emergence when trying to reconcile
theoretical, experimental, and computational predictions. 
Reference will 
be made to seminal papers and to
exhaustive reviews only. It is of course easy to wander a bit on arXiv and
look around the names of Calabrese, Pelissetto and Vicari, 
Prudnikov, Shalaev or Sokolov, Folk or Holovatch.
We apologise to those whose work is not directly mentioned in the short 
reference list, an indelicacy only due to our ignorance, our
misunderstanding or our laziness -- or all together.

\section{RG calculation of critical exponents}
Long distance properties of the Ising model near its second-order 
phase transition are described
in field theory by an effective Ginzburg-Landau-Wilson Hamiltonian
\be	H_{\rm Ising}[\varphi]=
	\int d^D\r\left[\undemi(\bnabla\varphi(\r))^2+\undemi
	m_0^2\varphi^2(\r)+{\textstyle\frac{\tilde u_0}{4!}}
	\varphi^4(\r)\right],
\label{eq-HamIM}
\ee
where $m_0^2$ is the bare coupling proportional to the 
deviation $T-T_c$ from the critical point
and $\varphi(\r)$ is a bare scalar field. Quenched 
randomness is introduced in such a model by
considering that the adjunction of disorder results in a 
distribution of local transition 
temperatures, so that a random temperature-like variable 
$\Delta$ is simply added to $m_0^2$,
\be	H[\varphi,\Delta]=\int d^D\r\left[\undemi(\bnabla\varphi(\r))^2+\undemi
	(m_0^2+\Delta)\varphi^2(\r)+
	{\textstyle\frac{\tilde u_0}{4!}}\varphi^4(\r)\right],
\label{eq-HamRIM}
\ee
where $\Delta$ is drawn from, e.g., a Gaussian probability 
distribution of zero mean and
dispersion $\sigma^2$, 
$\P(\Delta)=(2\pi\sigma^2)^{-1}
\exp(-\Delta^2/2\sigma^2)$.
For a specific disorder realization $[\Delta]$, the partition 
function and the free energy read
as $Z[\Delta]=\int\D[\varphi]e^{-\beta H[\varphi,\Delta]}$ 
and $F[\Delta]=-\beta^{-1}\ln Z[\Delta]$.
Average over quenched disorder then requires to calculate 
quantities like $\overline{F[\Delta]}=
-\beta^{-1}\int\D[\Delta]\ln Z[\Delta]\P(\Delta)$. This is 
performed through the introduction of 
$n$ replicas of the model (labelled by $\alpha$). Averaging over quenched 
disorder one ends up with an effective Hamiltonian with cubic anisotropy 
where the replicas are coupled through a new parameter $v_0$
\be	H_{\rm replicas}[\varphi]=\!\int\! d^D\r\!\left[\undemi\sum_{\alpha=1}^n
	\left[(\bnabla\varphi_\alpha(\r))^2+
	m_0^2\varphi_\alpha^2(\r)\right]\!+
	\!{\textstyle\frac{u_0}{4!}}\sum_{\alpha=1}^n\varphi_\alpha^4(\r)
	\!+\!
	{\textstyle\frac{v_0}{4!}}\left(\sum_{\alpha=1}^n
	\varphi_\alpha^2(\r)\right)^2
	\right].
\label{eq-HamEffRIM}
\ee
Here the bare coupling $u_0$, proportional to $\tilde u_0$,  
is positive and the bare coupling 
$v_0$, proportional to $-\sigma^2$, is negative.  
The properties of the random Ising model are recovered 
while taking the limit $n\to 0$, $\ln Z=\lim_{n\to 0}(Z^n-1)/n$.
Under a change of length scale by a factor $\mu$, the field and couplings are
renormalised according to
\be 	\varphi=Z_\phi^{1/2}\phi,\ m_0^2=Z_{m^2}m^2,
	\ u_0=\mu^\epsilon\frac{Z_u}{Z_\phi^2}u,
	\ v_0=\mu^\epsilon\frac{Z_v}{Z_\phi^2}v,
\label{eq-Zfactors}
\ee
where $\epsilon=4-D$. The RG functions are defined by 
differentiation at fixed bare
parameters, 
\begin{eqnarray}
 &\displaystyle \beta_u(u,v)=\left.\frac{\partial u}{\partial \ln\mu}\right|_0, 
 &\ \beta_v(u,v)=\left.\frac{\partial v}{\partial \ln\mu}\right|_0, 
	\\
 &\displaystyle \gamma_\phi(u,v)=\left.\frac{\partial 
	\ln Z_\phi}{\partial \ln\mu}\right|_0, 
 &\ \gamma_{m^2}(u,v)=\left.\frac{\partial \ln Z_{m^2}}
	{\partial \ln\mu}\right|_0.
\label{eq-betafunctions}
\end{eqnarray}
The skill of the theoretician is measured as his 
ability to compute these functions perturbatively, 
disentangling Feynman loops (they are known
up to 6 loops),
removing divergences which occur in the asymptotic 
limit by  controlled rearrangement of 
the series for the vertex functions. Eventually expecting reliable results 
after complicated resummation procedures~\cite{Yurko_etal}. 
Fixed points are then solutions of 
$\beta_u(u^*,v^*)=\beta_v(u^*,v^*)=0$, the stability 
of which is controlled by a stability matrix
$\frac{\partial\beta_i}{\partial u_j}$ with eigenvalues 
which besides the standard critical exponents also govern the corrections to
scaling (exponent $\omega$). 

At that point, even Simplicio may read off the 
critical exponents! Consider for example the pair correlation function of 
bare fields
$\<\varphi(0)\varphi(\r)\>$.
Under a change of length scale $\mu$, it renormalises to
$Z_\phi(\mu)\<\phi(0)\phi(\r)\>$. In the
same manner, for another dilatation parameter, $\mu s$,  one has
$\<\varphi(0)\varphi(s\r)\>\to Z_\phi(\mu s)\<\phi(0)\phi(s\r)\>$. 
The ratio from this latter to the
previous expression leads to
\be
	\frac{\<\phi(0)\phi(s\r)\>}{\<\phi(0)\phi(\r)\>}
	=\frac{Z_\phi(\mu)}{Z_\phi(\mu s)} 
	\frac{\<\varphi(0)\varphi(s\r)\>}{\<\varphi(0)\varphi(\r)\>}.
\ee
This expression gives the algebraic decay of the two-point 
correlation function of renormalised
fields $\<\phi(0)\phi(\r)\>\sim|\r|^{-(D-2+\eta_\phi)}$ 
in terms of the pair correlation
function of the bare fields which are described by mean-field theory (MFT),
i.e., at the Gaussian fixed point (FP),
$\<\varphi(0)\varphi(\r)\>\sim|\r|^{-(D-2)}$ ($\eta_{\rm MFT}=0$).
The ratio $\frac{Z_\phi(\mu)}{Z_\phi(\mu s)}=
e^{\int^\mu_{\mu s}\gamma_\phi d\ln\mu}$
evaluated at the new FP gives $s^{-\gamma_\phi^*}$ and leads to
\be
	\frac{\<\phi(0)\phi(s\r)\>}{\<\phi(0)\phi(\r)\>}
	\sim s^{-(D-2+\gamma_\phi^*)},
\ee
from which one reads off the value of the critical exponent at this FP:
\be\eta_\phi=\gamma_\phi^*.\ee 
Following the same argument, the scaling dimension $1/\nu$
of the (renormalised) temperature field $m^2$ is given at the random 
fixed point in terms of the MFT value,
$1/\nu_{\rm MFT}=2$, and one gets 
\be\frac 1\nu=2-\gamma_{m^2}^*.\ee
From these two exponents, the others may be deduced by scaling arguments, 
describing the leading singularities of
the physical quantities, e.g., of the magnetic susceptibility:
\be \chi(\tau)\sim \Gamma_\pm|\tau|^{-\gamma}, \quad 
\gamma=\nu(2-\eta_\phi).\label{eq-chi}\ee

%%% figure %%%
\begin{figure}[th]
  \centering\vskip-1cm
  \begin{minipage}{\textwidth}
  \epsfig{file=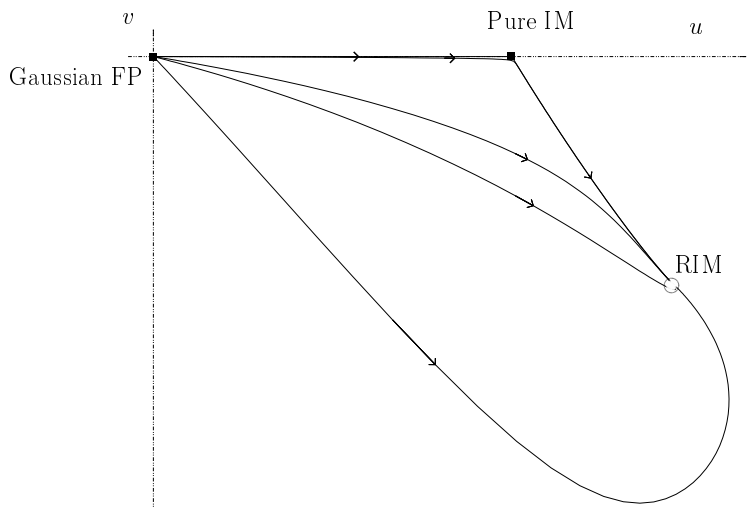,width=0.55\textwidth}%\qquad
  \end{minipage}\vskip-3cm
  \caption{RG flows in the $(u,v)$ plane. The points denoted ``Gaussian FP'', 
  ``Pure IM'' and ``RIM'' are respectively the Gaussian fixed point, the 
  pure Ising model fixed point and the random Ising model fixed point.
	}
	\label{Fig0}

\end{figure}
%%% end figure %%%

In the non-asymptotic regime, the system approaches 
criticality in a more complex way
and this is where corrections to scaling appear,
\be 
	\chi(\tau)\sim \Gamma_\pm |\tau|^{-\gamma} 
	(1+{\Gamma'}_\pm |\tau|^{\nu\omega}+
	{\Gamma ''}_\pm|\tau|^{2\nu\omega}+\dots),
\label{eq-chiCorr}
\ee
where the scaling dimension $\omega$ corresponds to the negative of the leading
 irrelevant RG eigenvalue, $\omega=-|y_3|$, as it is usually denoted,
and the dots in Eq.~(\ref{eq-chiCorr})
stand for higher order irrelevant corrections. 
{\em Non solum} the critical exponents, {\em sed etiam} combinations
of critical amplitudes and correction-to-scaling exponents are
universal quantities. 

Also it is common practice, in order to describe the approach to 
criticality especially in experiments and
simulations, to introduce effective exponents through
\be \chi(\tau)\sim \Gamma_\pm |\tau|^{-\gamma_{\rm eff}}, 
	\quad \gamma_{\rm eff}=-\frac{d\ln\chi(\tau)}{d\ln|\tau|}.
\label{eq-chiEff}
\ee
These effective exponents may be calculated theoretically 
from the flow equations, e.g.,
\begin{eqnarray}
	\eta_{\rm eff}(l)&=&\gamma_\phi(u(l),v(l)), \\ 
	\frac 1{\nu_{\rm eff}(l)}&=&2-\gamma_{m^2}(u(l),v(l)).
\end{eqnarray}
The variation of effective exponents depends on the RG flow in the
parameter space as shown in Fig.~\ref{Fig0}.

In the case of Heisenberg-like ferromagnets, the experimental observation
of a maximum of the effective exponent $\gamma_{\rm eff}$ found a 
theoretical explanation
in terms of trajectories in the parameter space~\cite{Folk2}. The same
observation holds in the case of the random Ising model, but the 
critical exponents also change at the disorder fixed point 
in this latter case.

\section{Experiments}
Experiments on site-diluted three-dimensional Ising magnets are usually
performed on uniaxial disordered anti-ferromagnets such as Fe$_{1-x}$Zn$_x$F$_2$
or Mn$_{1-x}$Zn$_x$F$_2$.
The original aim was the study of the random-field behaviour when a uniform 
magnetic field is applied to such a disordered system. However, 
when the samples are of high quality (low mosaicity, high chemical 
homogeneity), also the behaviour in zero external magnetic 
field is accessible ($3D$ disordered Ising model universality class). 
Staggered susceptibility and correlation length are deduced from neutron
scattering experiments. 
The scattering intensity
$I({\bf q})$ is the Fourier transform
of the pair correlation function, where long-range
fluctuations produce an isotropic Lorentzian peak
centred at the superstructure
spot position  ${\bf q}_0$ with a peak intensity given by the
susceptibility and a width determined by the inverse correlation length, while
long-range order gives a background  proportional to the order
parameter squared:
\begin{equation}
        I({\bf q})=\langle m^2\rangle\delta({\bf q}-{\bf q}_0)
        +\frac{\chi}{1+\xi^2({\bf q}-{\bf q}_0)^2}.
\end{equation}
Fitting the Lorentzian at different temperatures eventually give access to
the critical exponents, critical amplitudes, and possibly the correction
to scaling.

\section{Monte Carlo simulations}
The majority of  numerical studies of the disordered Ising 
model were concerned
with site dilution. But we may also choose to model 
the disorder by bond dilution in
order to compare these two kinds of disorder and to verify that 
they indeed lead
to the same set of new critical exponents, as
expected theoretically by
universality. In our study we therefore considered the
{\em bond-diluted\/} Ising model in three dimensions whose Hamiltonian with 
uncorrelated quenched
random interactions can be written (in a Potts model normalisation) as
\begin{equation}
-\beta {\cal H}=\sum_{(i,j)}K_{ij}\delta_{\sigma_i,\sigma_j},
\end{equation}
where the spins take the values $\sigma_i=\pm 1$ and the sum goes over
all nearest-neighbour pairs $(i,j)$. The coupling strengths are allowed to take
two different values $K_{ij}= K \equiv J/k_BT$ and $0$
with probabilities $p$ and $1-p$, respectively,
\begin{equation}
{\cal P}[K_{ij}]=
	\prod_{(i,j)}P(K_{ij})=
	\prod_{(i,j)}[p\delta(K_{ij}-K)+(1-p)\delta(K_{ij})],
\end{equation}
$c=1-p$ being the concentration of missing bonds, which play the role of the
non-magnetic impurities.

The phase diagram and the critical properties at a few
selected dilutions were studied by large-scale Monte Carlo
simulations on simple cubic lattices with $V=L^3$ spins (up to $L=96$) and
periodic boundary conditions in the three space directions, using the
Swendsen-Wang cluster algorithm for updating the spins.
All physical quantities are averaged over 2\,000 -- 5\,000 disorder
realisations, indicated by a bar (e.g., $\bar \chi$ for the susceptibility).
Standard definitions were used, e.g., for a given disorder realisation,
the magnetisation is defined according to $m=\langle|\mu|\rangle$ where
$\langle\dots\rangle$ stands for the thermal average and
$\mu=(N_\uparrow-N_\downarrow)/(N_\uparrow+N_\downarrow)$
with $N_{\uparrow,\downarrow}$ counting the number of ``up'' and ``down''
spins.
The susceptibility follows from the fluctuation-dissipation relation,
$\chi=KV(\langle\mu^2\rangle-\langle|\mu|\rangle^2)$. 
The phase diagram is obtained by locating the maxima of the average
susceptibility $\bar\chi_L$ (a diverging quantity in the thermodynamic limit)
for increasing lattice sizes $L$ as a function of the coupling strength $K$.

%%% figure %%%
\begin{figure}[ht]
  \centering
  \begin{minipage}{\textwidth}
  \epsfig{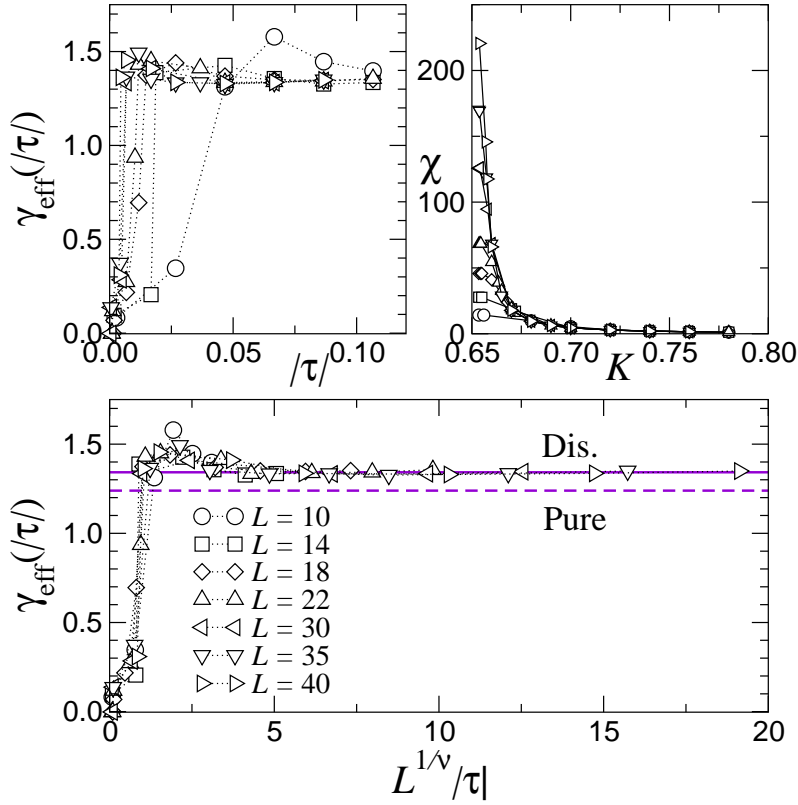}%\qquad
  \end{minipage}
  \caption{Variation of the temperature dependent effective critical exponent
	$\gamma_{\rm{eff}}(|\tau|)=-d\ln \bar\chi/d\ln |\tau|$
	(in the low-temperature phase)
	as a function of the reduced
	temperature $|\tau|$ (top) and $L^{1/\nu}|\tau|$ (bottom)
	for the bond-diluted Ising
	model with $p=0.7$ and several lattice sizes
	$L$.
	The horizontal solid and dashed lines indicate
	the site-diluted and pure values of $\gamma$, respectively.
	}
	\label{Fig1}
\end{figure}
%%% end figure %%%

As a function of the reduced temperature $\tau=(K_c- K)$ ($\tau<0$ in the
low-temperature (LT) phase and $\tau>0$ in the high-temperature (HT) phase)
and the system size $L$,
the  susceptibility is expected to scale as:
\begin{equation}
\bar\chi(\tau,L)\sim |\tau|^{-\gamma}g_\pm(L^{1/\nu}|\tau|),
\label{eq-chi-univ}
\end{equation}
where $g_\pm$ is a scaling function of the
variable $x=L^{1/\nu}|\tau|$ and the subscript $\pm$ stands 
for the HT/LT phases.
Recalling (\ref{eq-chiEff}) we can define a temperature dependent 
effective critical exponent
$\gamma_{\rm{eff}}(|\tau|)=-d\ln \bar\chi/d\ln |\tau|$,
which should converge towards
the asymptotic critical exponent $\gamma$ when
$L\rightarrow\infty$ and $|\tau|\rightarrow 0$. Our results for $p=0.7$
are shown in Fig.~\ref{Fig1}.
%%% figure %%%
\begin{figure}[ht]
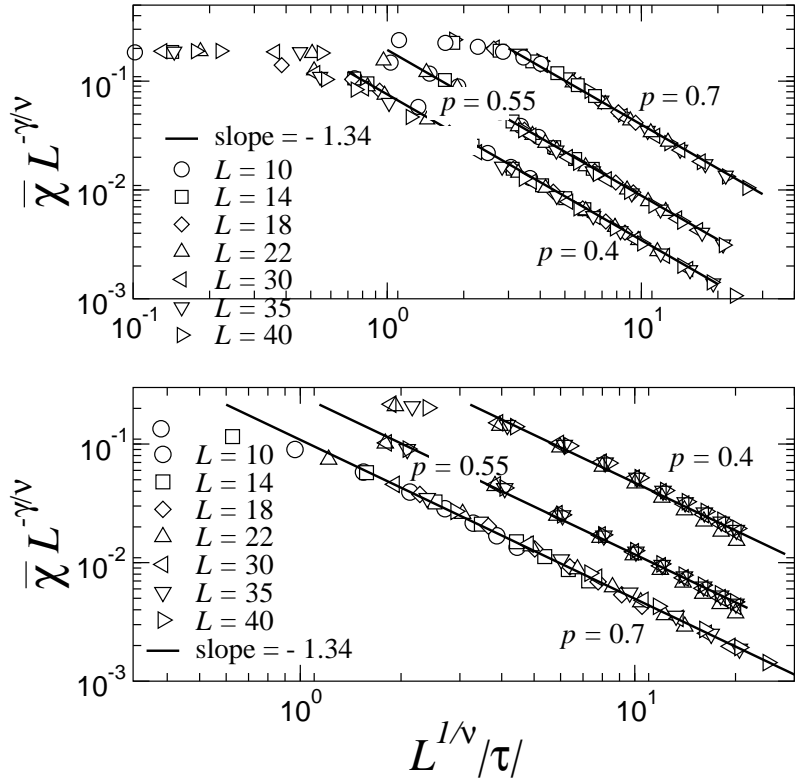

  \centering
  \begin{minipage}{\textwidth}
  \epsfig{file=scaling-chi-LT.eps,width=0.70\textwidth}
  \end{minipage}\vskip-3.0cm
  \begin{minipage}{\textwidth}
  \epsfig{file=scaling-chi-HT.eps,width=0.70\textwidth}
  \end{minipage}
  \caption{Log-log plot of the scaling function
$\bar\chi L^{-\gamma/\nu}$ in the LT and HT phases (respectively top and
bottom) against $L^{1/\nu}|\tau|$
for $p=0.4$, $0.55$, and
$0.7$. The solid lines
show the power-law behaviours with the exponent
$\gamma\simeq 1.34$ which characterise the disordered fixed point. The data for
the smallest values of $L^{1/\nu}|\tau|$,
which do not lye on the master curve, illustrate
the finite-size effects when the correlation length $\xi$ is
limited by the linear size $L$.
	}
	\label{Fig2}
\end{figure}
%%% end figure %%%
For the greatest sizes, the effective exponent 
$\gamma_{\rm{eff}}(|\tau|)$ is stable around $1.34$ when $|\tau|$ 
is not too small,
i.e., when the finite-size effects are not too strong. The plot
of $\gamma_{\rm{eff}}(|\tau|)$ vs.\ the rescaled variable $L^{1/\nu}|\tau|$
shows that
the critical power-law behaviour holds in different temperature ranges for the
different sizes studied. 
From the
temperature behaviour of the susceptibility, we also have directly extracted the
power-law exponent $\gamma$ using error weighted least-squares fits by
choosing the temperature range that gives the smallest $\chi^2/$d.o.f for
several system sizes. The results are
consistent with $\gamma \approx  1.34 - 1.36$.

From the previous expression of the susceptibility as a
function of the reduced temperature and size, it is instructive
to plot the scaling function $g_\pm(x)$.
For finite size and $|\tau|\not=0$, the scaling functions may be Taylor
expanded in powers of the inverse scaling variable
$x^{-1}=(L^{1/\nu}|\tau|)^{-1}$,
$\bar \chi_\pm(\tau,L)=|\tau|^{-\gamma} [g_\pm(\infty)+
x^{-1}g'_\pm(\infty)+O(x^{-2})]$,
where the amplitude $g_\pm(\infty)$ is usually denoted by $\Gamma_\pm$.
Multiplying by $L^{-\gamma/\nu}$ leads to
\begin{equation}
\bar\chi_\pm L^{-\gamma/\nu}=x^{-\gamma} g_\pm(x)=\Gamma_\pm 
x^{-\gamma}+O(x^{-\gamma-1}).
\end{equation}
The curves in the ordered and disordered phases,
shown in Fig.~\ref{Fig2},
are obviously universal master curves whose slopes, in a
log-log plot, give the critical exponent
$\gamma\simeq 1.34$. Indeed, when $|\tau|\rightarrow 0$ but with $L$ still
larger than the correlation length $\xi$, one should recover the critical
behaviour given by ${g}_\pm (x)=O(1)$.
The critical amplitudes $\Gamma_\pm$  follow.

\section{Results and conclusions}
In conclusion, according to Sagredo's suggestion
we may compare the results deduced from the different techniques. We only
concentrate here on the behaviour of the susceptibility, which  already
leads to partially conflicting results as can be seen by inspection of
Table~\ref{table}.
How to conclude in favour of universality? A possible answer would be the
following.
{%\small
\em\vskip2mm  \noindent In the same pub, a few days later, after
inspection of the results:
\begin{description}
\item{{\sc Sagr.}\ }Universality is still a good idea, but it is very
	difficult to produce high-quality samples where universality
	is clearly satisfied.
\item{{\sc Simpl.}\ }For the simulations, it is so time 
consuming to increase the size of
the system that it is not feasible at the moment. The problems here come
essentially from the thermodynamic limit and the disorder average.
\item{{\sc Salv.}\ }In the equations, the sample is perfect, the disorder
average is exact and the
thermodynamic limit is automatically
understood. So I believe that theoretical results
are correct.
\item{{\sc Simpl. and Sagr.}\ }But what about RG  calculations at 7
loop approximation? Will it come from St-Petersburg, from Roma, or from the 
Linz-Lviv axis?
\end{description}
}

\begin{table}[h]
\small
\caption{Critical exponents and critical amplitude ratio of the
susceptibility as measured with different techniques.}
\vglue0mm\begin{center}
        \begin{tabular}{@{}*{5}{l}}
\br
Technique & $\gamma$ & $\Gamma_+/\Gamma_-$ & $\omega$ & Ref. \\
\br
RG & &$2.2$ &  &     \cite{Newlove}\\
& $1.318$ & & $0.39(4)$ & \cite{Folk98,Folk00}$^1$ \\
& $1.330(17)$ & & $0.25(10)$ & \cite{Pelissetto1}$^2$ \\
\hline
Neutron scattering & $1.44(6)$ & $2.2$ & $0.5$ & \cite{Birgeneau0}$^3$ \\
	& $1.31(3)$ & $2.8(2)$ & & \cite{Belanger1}$^4$ \\
	& $1.37(4)$ & $2.40(2)$ & & \cite{Birgeneau1}$^5$\\
\hline
MC &	$1.342(10)$ & & $0.37$ &\cite{Ballesteros}$^6$ \\
& $1.34(1)$ & $1.62(10)$ & undetermined &\cite{Berche}$^7$ \\
&$1.342(7)$ & &undetermined&\cite{Calabrese1}$^8$ \\
\br
\centre{5}{\footnotesize $^1$ 4 loop approximation.}\\
\centre{5}{\footnotesize $^2$ 6 loop approximation,
	fixed dimension.}\\
\centre{5}{\footnotesize $^3$ Fe$_{1-x}$Zn$_x$F$_2$, $x=0.4$, 
	0.5, $|\tau|\sim 10^{-2}$.}\\
\centre{5}{\footnotesize $^4$ Fe$_{\rm 0.46}$Zn$_{\rm 0.54}$F$_2$, 
	$1.5 \times 10^{-3}\le|\tau|\le 10^{-1}$.}\\
\centre{5}{\footnotesize $^5$ Mn$_{\rm 0.75}$Zn$_{\rm 0.25}$F$_2$, 
	$4 \times 10^{-4}\le|\tau|\le 2 \times 10^{-1}$.}\\
\centre{5}{\footnotesize $^6$ site dilution, $p=0.4$ to 0.8.}\\
\centre{5}{\footnotesize $^7$ bond dilution, $p=0.7$.
	The correction to scaling is too small to be determined.}\\
\centre{5}{\footnotesize $^8$ site dilution, $p=0.8$. The observed 
	correction to scaling could be the 
	next-to-leading.}\\
\end{tabular}
\\
\medskip\end{center}
\label{table}
\end{table}

\noindent {\large\sf\fontseries{bx}\selectfont Acknowledgment }\par
It is a great pleasure to thank Yurko Holovatch who gave us the opportunity 
to contribute to the Festschrift dedicated to the 60th birthday of Reinhard Folk. 

%%%%%%%%%%%%%%%%%%%%%%%%%%%%%%%%%%%%%%%%%%%%%%%%%%%%%%%%%%%
%        REFERENCES
%%%%%%%%%%%%%%%%%%%%%%%%%%%%%%%%%%%%%%%%%%%%%%%%%%%%%%%%%%

%
%% If you have problems with typesetting in ukrainian uncomment line below.
%
  \end{document}